\documentstyle[preprint,epsfig,aps]{revtex}
\def\singlespace {\smallskipamount=3.75pt plus1pt minus1pt
                  \medskipamount=7.5pt plus2pt minus2pt
                  \bigskipamount=15pt plus4pt minus4pt
                  \normalbaselineskip=15pt plus0pt minus0pt
                  \normallineskip=1pt
                  \normallineskiplimit=0pt
                  \jot=3.75pt
                  {\def\smallskip {\vskip\smallskipamount}}
                  {\def\medskip   {\vskip\medskipamount}}
                  {\def\bigskip   {\vskip\bigskipamount}}
                  {\setbox\strutbox=\hbox{\vrule
                    height10.5pt depth4.5pt width 0pt}}
                  \parskip 4.5pt
                  \normalbaselines}
\def\middlespace {\smallskipamount=5.825pt plus1.5pt minus1.5pt
                  \medskipamount=11.25pt plus3pt minus3pt
                  \bigskipamount=22.5pt plus6pt minus6pt
                  \normalbaselineskip=22.5pt plus0pt minus0pt
                  \normallineskip=1pt
                  \normallineskiplimit=0pt
                  \jot=5.825pt
                  {\def\smallskip {\vskip\smallskipamount}}
                  {\def\medskip   {\vskip\medskipamount}}
                  {\def\bigskip   {\vskip\bigskipamount}}
                  {\setbox\strutbox=\hbox{\vrule
                    height15.75pt depth6.75pt width 0pt}}
                  \parskip 7.25pt
                  \normalbaselines}
\def\pr{\prime}
\def\be{\begin{equation}}
\def\lan{\left\langle}
\def\ran{\right\rangle}
\def\ee{\end{equation}}
\def\barr{\begin{array}}
\def\earr{\end{array}}

\def\nn8{\nonumber\\[10pt]}
\def\l{\left}
\def\r{\right}
\def\dis{\displaystyle}
\def\ed{\end{document}}
\def\fr{\{1^r\}}

\def\fnu{\{2^\nu 1^{N-2\nu}\}}
\def\fnr{\{1^{N-r}\}}

\begin{document}

\title{SU(N) Wigner-Racah algebra for the matrix of second moments of 
embedded Gaussian unitary ensemble of \\ random matrices}

\author{V. K. B.  Kota}
\address{Physical Research Laboratory, Ahmedabad 380 009, India}

\maketitle

\begin{abstract}

Recently Pluhar and Weidenm\"{u}ller [Ann. Phys. (N.Y.) {\bf 297}, 344
(2002)] showed that the eigenvectors of the matrix of second moments of
embedded  Gaussian unitary ensemble of random matrices generated by
$k$-body interactions (EGUE($k$)) for $m$ fermions in $N$ single particle
states are $SU(N)$ Wigner coefficients and derived also an expression
for the  eigenvalues. Going beyond this work, we will show that the
eigenvalues of this matrix are square of a $SU(N)$ Racah coefficient and
thus the matrix of second moments of EGUE($k$) is solved completely by
$SU(N)$ Wigner-Racah algebra.

\end{abstract}



\newpage
\narrowtext

\section{INTRODUCTION}

Interacting finite quantum systems such as nuclei, atoms, quantum dots,
nanometer-scale metallic grains etc. are governed by hamiltonians of low
(compared to the number of particles in the system) particle rank.
Therefore, for many purposes, the random matrix models appropriate for these
systems are embedded random matrix ensembles of $k$-body interactions
originally introduced, via nuclear shell model calculations, by French and
Wong and Bohigas and Flores \cite{Fr-71}. For a system of $m$ spinless
fermions in $N$ single particle states (we will use fermions throughout this
paper and turn briefly to bosons at  the end) the embedded  Gaussian unitary
ensemble of random matrices of $k$-body interactions [EGUE($k$)] is
generated by  defining the hamiltonian  $H$, which is given to be $k$-body, 
to be GUE in the $k$-particle spaces and then propagating it to the $m$
particle spaces by using the geometry (direct product structure) of the
$m$-particle spaces.  Just as EGUE($k$), the  EGOE($k$) and other embedded
ensembles are defined \cite{Ko-01}. With $m$ particle space dimension given
by $N_m=\l(^N_m\r)$,  one has the unitary groups $SU(N)$, $U(N_k)$ and
$U(N_m)$ with EGUE($k$) invariant under $U(N_k)$ and the embedding in
$m$-particle spaces is defined by $SU(N)$; note that a GUE in $m$ particle
spaces is invariant under $U(N_m)$ but not the EGUE($k$), $k < m$. Very
early,  using the so called binary correlation approximation, Mon and French
\cite{Mo-75} and later French et al \cite{Fr-88} derived some analytical
properties of embedded ensembles valid in the dilute limit (defined by 
$(N,m,k)  \rightarrow \infty$, $m/N \rightarrow 0$ and $k/m \rightarrow
0$).  However only recently rigorous analytical results,  valid for any
$(N,m,k)$ are  derived for these ensembles by Benet et al. \cite{Be-01} and
very soon Pluhar and Weidenm\"{u}ller (hereafter called PW) demonstrated
\cite{Pl-02} that these results indeed follow from considerations based on
the $SU(N)$ embedding algebra. With all the $m$-particle matrix elements
being linear combinations of the $k$-particle matrix elements (see Eq. (9)
ahead), the joint distribution for the matrix elements will be a
multivariate Gaussian. Thus all the information about EGUE($k$) is in the
covariance matrix or the matrix of second moments (Eq. (10) ahead). PW have
shown that the eigenvectors of this matrix are $SU(N)$ Wigner (or
Clebsch-Gordon (CG)) coefficients and derived the expression for their
eigenvalues using a duality relation for EGUE($k$). The purpose of this
paper is to show that the eigenvalues can be written as $SU(N)$ Racah
coefficients and thus the matrix of second moments is solved completely by
$SU(N)$ Wigner-Racah algebra. To this end results for $SU(N)$ Racah
coefficients given in \cite{He-75} and \cite{Bu-75,Bu-81} are used. We will
start with some basic results given in PW paper.

\section{Basic definitions and results}

Let us begin with $m$ particles in $N$ single particle states (unfortunately
in PW $\ell$ is used in place of $N$ but to keep the notations same as in our
earlier papers \cite{Ko-01,Ko-03}, we use $N$). The single particle (sp)
creation operator $a^\dagger_i$ for any $i$-th  sp state transforms as the
irreducible representation $\{1\}$ of $U(N)$ and similarly a product of $r$
creation operators transform, as we have fermions, as the  irrep  $\fr$ in
Young tableaux notation. Let us add that a $U(N)$ irrep $\{\lambda_1,
\lambda_2, \cdots, \lambda_N\}$ defines the corresponding $SU(N)$ irrep as
$\{\lambda_1-\lambda_N, \lambda_2-\lambda_N, \cdots, \lambda_{N-1}-
\lambda_N\}$ with $N-1$ rows (there are also other equivalent ways of
defining $SU(N)$ irreps given a $U(N)$ irrep \cite{Pl-02}). This $U(N)
\leftrightarrow SU(N)$ correspondence is used throughout and therefore we use
$U(N)$ and $SU(N)$ interchangeably. In PW, $\fr$ is denoted by $f_r$ and we
will follow this notation from now on.  With $v_r$ denoting irreps (and other
multiplicity labels) of the groups in a subgroup chain of $U(N)$ that supply
the labels needed for a complete specification of any $m$-particle state (for
the purpose of the present paper the subgroup chain need not be specified),
the operator $\dis\prod_{i=1}^r\, a^\dagger_i$ and a normalized $r$-particle
creation operator $A^\dagger(f_r v_r)$ behave as the $SU(N)$ tensors $T^{f_r
v_r}$ and $\frac{1}{\sqrt{r!}}T^{f_r v_r}$ respectively. Using the
composition formula,
\be
T^{f_m v_m} = \dis\sum_{v_k,v_s}\;C^{f_m v_m}_{f_k v_k\;f_s v_s} T^{f_k v_k} 
T^{f_s v_s}\;,\;\;s=m-k
\ee
where $C^{f_m v_m}_{f_k v_k\;f_s v_s}$ is a $SU(N)$ CG coefficient, a
$m$-particle state $\l.\l| f_m v_m\r.\ran = A^\dagger(f_m v_m) \l.\l| 0
\r.\ran$ can be written as a product of $k$ and $s=m-k$ particle states as,
\be
\l.\l| f_m v_m \r.\ran=\l(^m_k\r)^{-\frac{1}{2}}\; \dis\sum_{v_k
v_s} \;A^\dagger(f_k v_k) \l.\l|f_s v_s\r.\ran \,C^{f_m v_m}_{f_k v_k\;
f_s v_s}
\ee
Some properties of the CG coefficients, used later in simplifications, are
\be
\barr{l}
C^{f_{ab} v_{ab}}_{f_a v_a\;f_b v_b} = (-1)^{\phi(f_a, f_b, f_{ab})} \,
C^{f_{ab} v_{ab}}_{f_b v_b\;f_a v_a}\,,\;\;\;
C^{f_{ab} v_{ab}}_{f_a v_a\;f_b v_b} = C^{\overline{f_{ab}} \overline{v_{ab}
}}_{\overline{f_a} \overline{v_a}\;\overline{f_b} \overline{v_b}} \nn8
C^{f_{ab} v_{ab}}_{f_a v_a\;f_b v_b} = (-1)^{\phi(f_a, f_b, f_{ab})}\;
\dis\sqrt{\dis\frac{d(f_{ab})}{d(f_a)}}\;
C^{f_{a} v_{a}}_{f_{ab} v_{ab}\;\overline{f_b} \overline{v_b}} \nn8
C^{f_a v_a}_{00\;f_a v_a}=1,\;\; C^{00}_{f_a v_a\;\overline{f_a} \overline{
v_a}}=\dis\frac{1}{\sqrt{d(f_a)}}\,,\;\; 
\l(C^{f_{ab} v_{ab}}_{f_a v_a\;\overline{f_a} \overline{v_b}}\r)^* =
C^{f_{ab} v_{ab}}_{f_a v_b\;\overline{f_a} \overline{v_a}} \nn8
\dis\sum_{v_a\,,\, v_b} \l(C^{f_{ab} v_{ab}}_{f_a v_a\;\overline{f_a} 
\overline{v_b}}\r)^* \;C^{f^\pr_{ab} v^\pr_{ab}}_{f_a v_a\;\overline{f_a} 
\overline{v_b}} = \delta_{f_{ab} f^\pr_{ab}}\;\delta_{v_{ab} v^\pr_{ab}}
\earr
\ee
In (3), $\phi$ is a function that defines the phase for the  $1
\leftrightarrow 2$ interchange in the CG coefficients, $d(f)$ is the
dimension of the irrep $f$ and $\overline{f}$ is the irrep conjugate to
${f}$. For $f_r=\fr$, $\overline{f_r}=\fnr$  and it also contains a phase
factor as given in Eq. (53) of \cite{He-75} (this is also seen  easily by
comparing the second and third equalities in (3) with the  corresponding
relations for the standard CG coefficients for angular momentum 
\cite{Ed-74}). Similarly with $f_a=f_k$, $f_b=\overline{f_k}$, one has 
$$
f_{ab}=g_\nu =\fnu, \;\;\;\nu=0,1,\ldots,k.
$$
Note that $g_0=\{0\}$ for $SU(N)$ and also $g_\nu = \overline{g_\nu}$. The
function $\phi$ in Eq. (3) is of the form $\phi(\lambda_1, \lambda_2,
\lambda_3) = F(\lambda_1) + F(\lambda_2) + F(\lambda_3)$  where $F$ is some
function with $F(\lambda)$ an integer and $F(\lambda) =
F(\overline{\lambda})$; for $SU(N)$  irreps that appear in this paper these
results are valid as can be seen from Eq. (60) of \cite{He-75}) except that
there can be an overall $N$ dependent factor which will not change any of
the final results. 

$SU(N)$ irreducible tensors $B_k(g_\nu  \omega_\nu)$ constructed out of
$A^\dagger(f_k v_k)A(f_k v^\pr_k)$ are defined by,
\be
B_k(g_\nu \omega_\nu) = \dis\sum_{v_k, v_k^\prime}\,A^\dagger(f_k v_k) 
A(f_k v^\prime_k)\;C^{g_\nu \omega_\nu}_{f_k v_k\;\overline{f_k} 
\overline{v^\prime_k}} 
\ee
It is useful to note that the tensors $B$'s in (4) multiplied by $k!$ are, 
to within a phase factor, same as the tensors defined in Eq. (48) of 
\cite{He-75}. Wigner-Eckart theorem decomposes the matrix elements of 
$B_k(g_\nu \omega_\nu)$ in $m$-particle spaces into a reduced matrix element
$\lan\;\mid\mid\;\;\;\mid\mid\;\ran$ and a CG coefficient,
\be
\lan f_m v_m \mid B_k(g_\nu \omega_\nu) \mid f_m v^\prime_m\ran = \lan f_m 
\mid\mid B_k(g_\nu) \mid\mid f_m\ran\;C^{g_\nu \omega_\nu}_{f_m v_m\;
\overline{f_m} \overline{v^\prime_m}}
\ee
Two important properties of $B_k(g_\nu \omega_\nu)$ are,
\be
\barr{l}
\lan\lan B_k(g_\nu \omega_\nu) \ran\ran^k = \dis\sqrt{N_k}\,
\delta_{g_\nu,\{0\}} \nn8
\lan\lan B_k(g_\nu \omega_\nu) B_k(g_\mu \omega_\mu)
\ran\ran^k =\delta_{g_\nu g_\mu}\;\delta_{\omega_\nu 
\omega_\mu}
\earr
\ee 
In (6), $\lan\lan\;\ran\ran^k$ denotes trace over the $k$-particle spaces.
The first equality  in Eq. (6) easily follows from the fact that here
$g_\nu=\{0\}$ as traces  are scalars with respect to $SU(N)$ and then
applying the fifth equality in  Eq. (3). Similarly the second equality in
Eq. (6) follows from the fact that only $g_\nu=g_\mu$ will give a scalar.
Now we will turn to EGOE($k$). 

Consider a $k$-body hamiltonian 
\be
H(k) = \dis\sum_{v_a,\, v_b}\; V_{v_a v_b}(k) A^\dagger(f_k v_a) A(f_k v_b)
\ee
where $V_{v_a v_b}(k)$ are matrix elements of $H(k)$ in $k$-particle space
and form a GUE, i.e. $V_{v_a v_b}(k)$ are independent Gaussian variables
with zero center and variance given by 
\be
\overline{V_{v_a v_b}(k)\;V_{v_c v_d}(k)} = \dis\frac{\lambda^2}{N_k}\;
\delta_{v_a v_d} \delta_{v_b v_c}
\ee
In Eq. (8) the overline indicates ensemble average and $\lambda^2$ is 
ensemble averaged variance of $H(k)$ in $k$-particle space. The 
$m$-particle matrix elements of $H(k)$ are, with $s=m-k$,
\be
H_{v_m^1 v_m^2}(k) = \lan f_m v_m^1 \mid H(k) \mid f_m v_m^2\ran =
\l(^m_k\r)\;\dis\sum_{v_a, v_b, v_s} \l({C^{f_m v_m^1}_{f_k v_a f_s v_s}
}\r)^* \;C^{f_m v_m^2}_{f_k v_b f_s v_s}\;V_{v_a v_b}(k)
\ee
Eq. (9) is obtained easily by substituting the definition (7) for $H(k)$,
then inserting complete set of states between $A^\dagger$ and $A$ operators
and applying Eq. (2).  The EGUE($k$) in $m \geq k$ spaces is defined by Eqs.
(7)-(9). Now it is clear that for any analysis of EGUE($k$) all one need to
know is the covariance between any two $m$-particle matrix elements
$H_{v_m^1 v_m^2}(k)$ and this defines the matrix of second moments,
\be
A_{v_m^1 v_m^4\,:\,v_m^3 v_m^2} = \overline{H_{v_m^1 v_m^2}(k) H_{v_m^3
v_m^4}(k)}
\ee
As stressed by PW, most important step in EGUE($k$) analysis is to derive a
"generalized eigenvalue expansion" of $A$ defined by $A_{ij}=\sum_k\;C_{ik}
E_k C_{jk}$ with $E_k$ the eigenvalues and  $C_{jk}$ the eigenvectors such
that $E_k$ are positive and $C$'s  hermitian. To this end, it is useful to
consider the unitary decomposition  of $H(k)$ in terms of the $SU(N)$
tensors $B_k(g_\nu \omega_\nu)$,
\be
H(k) = \dis\sum_{g_\nu, \omega_\nu}\;B_k(g_\nu \omega_\nu)\;W_{g_\nu
\omega_\nu}(k)
\ee
The expansion coefficients $W_{g_\nu \omega_\nu}(k)$ are easily given by
\be
W_{g_\nu \omega_\nu}(k) = <<H(k) B_k(g_\nu \omega_\nu)>>^k
\ee
and this follows by using the definition (11) and Eq. (6). Most significant
property of the $W$ coefficients is that they are independent Gaussian
variables with zero center and variance given by (derived using Eqs. (11),
(4), (7) and (8) in that order and using the orthonormal properties of the 
CG coefficients), 
\be
\overline{W_{g_\nu \omega_\nu}(k) W_{g_{\mu} \omega_{\mu}}(k)} =
\dis\frac{\lambda^2}{N_k}\;\delta_{g_\nu g_{\mu}}\;\delta_{\omega_\nu
\omega_{\mu}}
\ee

\section{Matrix of second moments}

Firstly we will derive an expression for the covariance
$\overline{H_{v_m^1  v_m^2}(k) H_{v_m^3 v_m^4}(k)}$ in terms of $SU(N)$ CG
coefficients and Racah coefficients and then turn to the eigenvalues and
eigenvectors of $A$, the matrix of second moments. Applying Eqs. (10), (11)
and (5) in that order gives,
\be
\barr{l}
\overline{H_{v_m^1 v_m^2}(k) H_{v_m^3 v_m^4}(k)} \nn8
= \dis\sum_{g_\nu,\omega_\nu,g_\mu,\omega_{\mu}}\;\overline{
\lan f_m v_m^1 \mid B_k(g_\nu \omega_\nu)\;W_{g_\nu \omega_\nu}(k) \mid 
f_m v_m^2\ran \lan f_m v_m^3 \mid B_k(g_{\mu} \omega_{\mu})\;
W_{g_{\mu} \omega_{\mu}}(k) \mid f_m v_m^4\ran} \nn8
= \dis\sum_{g_\nu,\omega_\nu,g_{\mu},\omega_{\mu}}\;
\overline{W_{g_\nu \omega_\nu}(k) W_{g_{\mu} \omega_{\mu}}(k)}\;
\lan f_m v_m^1 \mid B_k(g_\nu \omega_\nu) \mid f_m
v_m^2\ran \lan f_m v_m^3 \mid B_k(g_{\mu} \omega_{\mu}) \mid 
f_m v_m^4\ran \nn8
=\dis\frac{\lambda^2}{N_k}\;\dis\sum_{g_\nu \omega_\nu}\;
\lan f_m v_m^1 \mid B_k(g_\nu \omega_\nu) \mid f_m
v_m^2\ran \lan f_m v_m^3 \mid B_k(g_\nu \omega_\nu) \mid 
f_m v_m^4\ran \nn8
= \dis\frac{\lambda^2}{N_k}\;\dis\sum_{g_\nu \omega_\nu}\;
\l|\lan f_m \mid\mid B_k(g_\nu) \mid\mid f_m\ran\r|^2\;
C^{g_\nu \omega_\nu}_{f_m v_m^1 \overline{f_m} \overline{v_m^2}}\;
C^{g_\nu \omega_\nu}_{f_m v_m^3 \overline{f_m} \overline{v_m^4}}
\earr
\ee
Eqs. (55) and (56) of \cite{He-75} together with Eqs. (3)  
(see also the remark just after Eq. (4)) allows one to write the reduced 
matrix element in Eq. (14) as a $SU(N)$ Racah or $U$-coefficient,
\be
\l|\lan f_m \mid\mid B_k(g_\nu) \mid\mid f_m\ran\r|^2 =
\dis\frac{(N_m)^2 \l(^m_k\r)^2}{d(g_\nu) (N_{m-k})}\;\l[
U(f_m f_{N-k} f_m f_k ; f_{m-k} g_\nu)\r]^2
\ee
With $\{2^\nu 1^{N-2\nu}\} = \{1^\nu\} \otimes \{1^{N-\nu}\} -
\{1^{\nu-1}\}  \otimes \{1^{N-\nu+1}\}$ where $\otimes$ denotes Kronecker
product, the dimension $d(g_\nu)$ is given by
\be
d(g_\nu) = d(\nu) = (N_\nu)^2 - (N_{\nu-1})^2 = \dis\frac{(N!)^2 (N+1)
(N-2\nu+1)}{(\nu!)^2 (N-\nu+1!)^2}
\ee
Before going further let us define, for a given $(m,N)$, a function 
$\Lambda^\nu(k)$,
\be
\Lambda^\nu(k) = {\tiny{\l(\barr{c} m-\nu \\ k \earr\r)}}\;{\tiny{\l(
\barr{c} N-m+k-\nu \\ k \earr\r)}}
\ee
Now, using Eq. (16) and substituting the expression given by Eq. (61) of 
\cite{He-75} for the $U$-coefficient in (15), it is seen that
\be
\l|\lan f_m \mid\mid B_k(g_\nu) \mid\mid f_m\ran\r|^2 =
\Lambda^\nu(m-k)\;,\;\;\;\nu=0,1,2,\ldots,k
\ee
Combining (14) with (18) yield an expression for the covariance between $H$
matrix elements in $m$-particle spaces,
\be
\barr{l}
\overline{H_{v_m^1 v_m^2}(k) H_{v_m^3 v_m^4}(k)} \nn8
=\dis\frac{\lambda^2}{N_k}\;\dis\sum_{\nu=0,1,\ldots,k;\, \omega_\nu}\;
\dis\frac{(N_m)^2 \l(^m_k\r)^2}{d(g_\nu) (N_{m-k})}\;\l[
U(f_m f_{N-k} f_m f_k ; f_{m-k} g_\nu)\r]^2\;
C^{g_\nu \omega_\nu}_{f_m v_m^1 \overline{f_m} \overline{v_m^2}}\;
C^{g_\nu \omega_\nu}_{f_m v_m^3 \overline{f_m} \overline{v_m^4}} \nn8
=\dis\frac{\lambda^2}{N_k}\;\dis\sum_{\nu=0,1,\ldots,k;\, \omega_\nu}\;
\l\{\Lambda^\nu(m-k)\r\}\;
C^{g_\nu \omega_\nu}_{f_m v_m^1 \overline{f_m} \overline{v_m^2}}\;
C^{g_\nu \omega_\nu}_{f_m v_m^3 \overline{f_m} \overline{v_m^4}}
\earr
\ee
Eq. (19) will be useful in deriving expressions for the moments of $H$
spectrum, i.e. $\overline{\lan H^p\ran^m}$. However the disadvantage of
(19) is that it is not in a proper form to give the eigenvalues and
eigenvectors of the matrix $A$ in Eq. (10). In order obtain them, the  CG
coefficients in  Eq. (14) should be changed to $C^{g_{\mu}
\omega_{\mu}}_{f_m v_m^1  \overline{f_m} \overline{v_m^4}}\;C^{g_{\mu}
\omega_{\mu}}_{f_m v_m^3  \overline{f_m} \overline{v_m^2}}$ and this can be
accomplished by a $SU(N)$ Racah transform. Using Eq. (3.2.17) of
\cite{Bu-81} one has for example,
\be
\dis\sum_{\omega_\nu}\; C^{g_\nu \omega_\nu}_{f_m v_m^1 \overline{f_m} 
\overline{v_m^2}}\; C^{f_m v_m^4}_{g_\nu \omega_\nu f_m v_m^3} =
\dis\sum_{g_{\mu} \omega_{\mu}}\;U(f_m \overline{f_m} f_m f_m; g_\nu
g_{\mu})\; C^{g_{\mu} \omega_{\mu}}_{\overline{f_m} 
\overline{v_m^2} f_m v_m^3}\; C^{f_m v_m^4}_{f_m v_m^1 g_{\mu} \omega_{
\mu}}
\ee
Then Eqs. (20) and (3) will give,
\be
\barr{l}
\dis\sum_{\omega_\nu}\; C^{g_\nu \omega_\nu}_{f_m v_m^1 \overline{f_m} 
\overline{v_m^2}}\;
C^{g_\nu \omega_\nu}_{f_m v_m^3 \overline{f_m} \overline{v_m^4}} \nn8
= \dis\sum_{g_{\mu}
\omega_{\mu}}\;\dis\sqrt{\dis\frac{d(g_\nu)}{d(g_{\mu})}}
\;U(f_m \overline{f_m} f_m f_m; g_\nu g_{\mu})
C^{g_\mu \omega_\mu}_{f_m v_m^1 \overline{f_m} \overline{v_m^4}}\;
C^{g_{\mu} \omega_{\mu}}_{f_m v_m^3 \overline{f_m} \overline{v_m^2}} 
\earr
\ee
Finally Eqs. (21) and (15) combined with (14) produce the generalized
eigenvalue expansion of the matrix of second moments $A$,
\be
\barr{l}
A_{v_m^1 v_m^4\,:\,v_m^3 v_m^2} = 
\overline{H_{v_m^1 v_m^2}(k) H_{v_m^3 v_m^4}(k)} = 
\dis\sum_{g_{\mu} \omega_{\mu}}\;
C^{g_\mu \omega_\mu}_{f_m v_m^1 \overline{f_m} \overline{v_m^4}}\;
C^{g_{\mu} \omega_{\mu}}_{f_m v_m^3 \overline{f_m} \overline{v_m^2}} 
\;\l\{\dis\frac{\lambda^2}{N_k}\;
\dis\frac{(N_m)^2 \l(^m_k\r)^2}{(N_{m-k})}\r. \nn8
\l.\l[\dis\sum_{g_\nu} \dis\sqrt{\dis\frac{1}{d(g_\nu) d(g_{\mu})}}\;
\l[U(f_m f_{N-k} f_m f_k ; f_{m-k} g_\nu)\r]^2\;
U(f_m \overline{f_m} f_m f_m; g_\nu g_{\mu})\r]\r\}
\earr
\ee
Obviously the quantity in the curly brackets in Eq. (22) gives the 
eigenvalues of $A$ and the $C$'s are eigenvectors. 

\section{Eigenvalues as $SU(N)$ Racah coefficients and their applications} 

In order to proceed further, it is useful to consider $6j$ symbols of $SU(N)$ 
and they are defined by (see Eq. (3.2.18) of \cite{Bu-81}),
\be
U(\lambda_1 \lambda_2 \lambda \lambda_3\, ;\,\lambda_{12} \lambda_{23}) 
= \dis\sqrt{d(\lambda_{12}) d(\lambda_{23})} (-1)^{\phi(\lambda_2,
\overline{\lambda_2}, {0}) + \phi(\lambda_{12}, \lambda_3,
\overline{\lambda}) + \phi(\lambda_1, \lambda_2,
\overline{\lambda_{12}})}\;
\l\{\barr{ccc} \lambda_1 & \lambda_{23} & \overline{\lambda} \\
\overline{\lambda_3} & \lambda_{12} & \lambda_2 \earr \r\}
\ee
In (23) $\lambda$'s are $SU(N)$ irreps and the four couplings involved  in
the $U$-coefficient are assumed to be multiplicity free (for the
applications in the present paper this assumption is always valid). 
Symmetry properties of the $6j$-symbol appearing on the r.h.s of (23) are
well known \cite{Bu-75,Bu-81}. In the present analysis, the
Biedenharn-Elliott sum rule  extended to  $SU(N)$ \cite{Bu-75,Bu-81} plays a
central role. This sum rule relates a product of three Racah coefficients
(weighted appropriately by dimension factors and phase factors with the
irreps in the Racah coefficients appearing in some particular order) with
sum over a common irrep label to a product of two Racah coefficients.
After converting the Racah coefficients in (22) into $6j$ symbols of $SU(N)$
using Eq. (23) and then applying the symmetry properties of the $6j$
symbols, it is seen that the sum in the square  brackets in Eq. (22) is
exactly in the required form. Applying the Biedenharn-Elliott sum rule, the
sum then simplifies to
$$
\dis\frac{N_{m-k}}{N_k \; d(g_{\mu})}\;
U^2(f_m f_{N-m+k} f_m f_{m-k}\,;\, f_k g_{\mu})
$$
Here $\mu=0,1,\ldots,m-k$. Now the eigenvalues of the matrix $A$, in
terms of the $U(N)$ Racah coefficients is given by
\be
E_{\mu} = \dis\frac{\lambda^2}{N_k}\;
\dis\frac{(N_m)^2 \l(^m_k\r)^2}{d(g_\mu) (N_k)}\;
\l[U(f_m f_{N-m+k} f_m f_{m-k}; f_k g_{\mu})\r]^2\,;\;\; \mu=0,1,\cdots,m-k
\ee
with degeneracy $d(g_{\mu})$ (see Eq. (16)). Eq. (24) is the central result
of this paper. With this, the matrix $A$ is completely specified by the
$U(N)$ Wigner and Racah coefficients. Now substituting the formula (Eq.
(61)) of \cite{He-75}) for the $U$-coefficients in Eq. (24) produces the
result of PW,
\be
A_{v_m^1 v_m^4\,:\,v_m^3 v_m^2} = 
\overline{H_{v_m^1 v_m^2}(k) H_{v_m^3 v_m^4}(k)} = 
\dis\sum_{g_{\mu} \omega_{\mu}}\;
C^{g_\mu \omega_\mu}_{f_m v_m^1 \overline{f_m} \overline{v_m^4}}\;
C^{g_{\mu} \omega_{\mu}}_{f_m v_m^3 \overline{f_m} \overline{v_m^2}} 
\;E_{\mu}
\ee
where, 
\be
E_{\mu} = \dis\frac{\lambda^2}{N_k}\;\Lambda^{\mu}(k)\,,\;\;\;\;\mu=0,1,
\cdots,m-k.
\ee
Note that the function $\Lambda^\mu(k)$ is defined by Eq. (17).

Information about EGUE($k$) is contained in the ensemble averaged moments 
$M_p=\overline{\lan H^p\ran^m}$ and the bivariate moments $\Sigma_{pq} = 
\overline{\lan H^p\ran^m \; \lan H^q\ran^m}$. In deriving the formulas for
the lower order moments, we will show the usefulness of Eq. (19). 
Obviously, ensemble averaged centroid is zero and the variance is
\be
\barr{rcl}
\overline{\lan H^2\ran^m} & = & \frac{1}{N_m}\;\dis\sum_{v_m^i,v_m^j}\;
\overline{H_{v_m^i v_m^j} H_{v_m^j v_m^i}} \nn8
& = & \frac{1}{N_m}\;\dis\sum_{g_\mu,\omega_\mu}\; \Lambda^\mu(k) 
\;\dis\sum_{v_m^i,v_m^j}\;
C^{g_\mu \omega_\mu}_{f_m v_m^i \overline{f_m} \overline{v_m^i}}\; 
C^{g_\mu \omega_\mu}_{f_m v_m^j \overline{f_m} \overline{v_m^j}} \nn8
& = & \Lambda^0(k)
\earr
\ee
The second equality follows from (25) and the final result follows by
applying (3); note that $\sum_{v_m^i}\;C^{g_\mu \omega_\mu}_{f_m v_m^i
\overline{f_m}  \overline{v_m^i}} =\sqrt{N_m}\,\delta_{\mu,0}$. 
The variance in (27) is in $\lambda^2/N_k$ units and this
factor is dropped as all the quantities we consider from now on are all 
scaled with respect to $\{\overline{\lan H^2 \ran^m}\}^{1/2}$. As the third
moment is zero, we will turn to the fourth moment,
\be
\barr{l}
\overline{\lan H^4\ran^m} = \dis\frac{1}{N_m}\;\dis\sum_{v_m^i,v_m^j,
v_m^{k^\pr},v_m^l}\;
\overline{ H_{v_m^i v_m^j} H_{v_m^j v_m^{k^\pr}} H_{v_m^{k^\pr} v_m^l} 
H_{v_m^l v_m^i}}
\nn8
= 
\dis\frac{1}{N_m}\;\dis\sum_{v_m^i,v_m^j,v_m^{k^\pr},v_m^l}\;
\l\{2\l[\dis\sum_{g_\nu,\omega_\nu}\;
\lan f_m v_m^i \mid B_k(g_\nu \omega_\nu) \mid f_m v_m^j\ran
\lan f_m v_m^j \mid B_k(g_\nu \omega_\nu) \mid f_m v_m^{k^\pr}\ran\r] \;
\times \r. \nn8
\l[\dis\sum_{g_\mu,\omega_\mu}\;
\lan f_m v_m^{k^\pr} \mid B_k(g_\mu \omega_\mu) \mid f_m v_m^l\ran
\lan f_m v_m^l \mid B_k(g_\mu \omega_\mu) \mid f_m v_m^i\ran\r] \nn8
+ \l[\dis\sum_{g_\nu,\omega_\nu}\;
\lan f_m v_m^i \mid B_k(g_\nu \omega_\nu) \mid f_m v_m^j\ran
\lan f_m v_m^{k^\pr} \mid B_k(g_\nu \omega_\nu) \mid f_m v_m^l\ran\r]
\;\times
\nn8
\l.\l[\dis\sum_{g_\mu,\omega_\mu}\;
\lan f_m v_m^j \mid B_k(g_\mu \omega_\mu) \mid f_m v_m^{k^\pr}\ran
\lan f_m v_m^l \mid B_k(g_\mu \omega_\mu) \mid f_m v_m^i\ran\r] \r\}\nn8
=2\l[\Lambda^0(k)\r]^2 + \dis\frac{1}{N_m}\;\dis\sum_{v_m^i,v_m^j,
v_m^{k^\pr},v_m^l} \l\{\dis\sum_{\nu=0,1,\ldots,k;\, \omega_\nu}\;
\l\{\Lambda^\nu(m-k)\r\}\;
C^{g_\nu \omega_\nu}_{f_m v_m^i \overline{f_m} \overline{v_m^j}}\;
C^{g_\nu \omega_\nu}_{f_m v_m^{k^\pr} \overline{f_m} \overline{v_m^l}}\r\}
\times \nn8
\l\{\dis\sum_{\mu=0,1,\ldots,m-k;\, \omega_\mu}\;
\l\{\Lambda^\mu(k)\r\}\;
C^{g_\mu \omega_\mu}_{f_m v_m^j \overline{f_m} \overline{v_m^i}}\;
C^{g_\mu \omega_\mu}_{f_m v_m^l \overline{f_m} \overline{v_m^{k^\pr}}}\r\}
\nn8
= 2\l[\Lambda^0(k)\r]^2 + \dis\frac{1}{N_m} \dis\sum_{\nu=0}^{min\{k,m-k\}}
\;\Lambda^\nu(m-k)\;\Lambda^\nu(k)\;d(\nu)
\earr
\ee
The second equality in Eq. (28) follows by applying Eqs. (11) and (13). In 
the third equality, it is easy to recognize the first term. The second term
follows by applying Eq. (19) and (25) to the two pieces in the 
corresponding term in the second equality. The final result follows by
applying the orthonormality of the  CG coefficients. Eqs. (27,28) will give
the excess ($\gamma_2$) parameter of the density of eigenvalues of
EGUE($k$),
\be
\gamma_2=\dis\frac{\overline{\lan H^4\ran^m}}{\l[\overline{\lan H^2\ran^m}
\r]^2}\;-\;3 = \l[\dis\frac{1}{N_m}\;\dis\sum_{\nu=0}^{min\{k,m-k\}}\; 
\dis\frac{\Lambda^\nu(m-k)\;\Lambda^\nu(k)\;d(\nu)}{\l[\Lambda^0(k)\r]^2}\r]
\;-1
\ee
Turning now to the lowest bivariate moment $\Sigma_{11}$, it is easily 
seen that
\be
\Sigma_{11} = \overline{\lan H \ran^m \lan H \ran^m} = \dis\frac{1}{
(N_m)^2}\;\dis\sum_{v_m^i,v_m^j} \overline{H_{v_m^i v_m^i} H_{v_m^j v_m^j}}
= \dis\frac{1}{N_m}\;\Lambda^0(m-k)
\ee
Applying (19) and recognizing that only $\nu=0$ will contribute to the
traces give immediately Eq. (30). However Eq. (25) generates a different
formula for $\Sigma_{11}$ and  equating it to (30) gives the identity
(derived in PW using the duality transformation),
$\dis\frac{1}{N_m}\;\sum_{\nu=0}^{m-k}\; \Lambda^\nu(k)  d(\nu) =
\Lambda^0(m-k)$. The variance of the distribution of centroids of  the $H$
spectra over the EGUE($k$) ensemble is 
\be
{\hat{\Sigma}}_{11} = \dis\frac{\Sigma_{11}}{\overline{\lan H^2\ran^m}}
=\dis\frac{1}{N_m}\;\dis\frac{\Lambda^0(m-k)}{\Lambda^0(k)}
\ee
Finally $\Sigma_{22}$ is given by
\be
\barr{l}
\Sigma_{22} =  \overline{\lan H^2 \ran^m \lan H^2 \ran^m} =
\dis\frac{1}{(N_m)^2}\;
\dis\sum_{v_m^i,v_m^j,v_m^{k^\pr},v_m^l} \overline{\l|H_{v_m^i v_m^j}\r|^2
\l|H_{v_m^{k^\pr} v_m^l}\r|^2} \nn8
= \dis\frac{1}{(N_m)^2}\;
\dis\sum_{v_m^i,v_m^j,v_m^{k^\pr},v_m^l} \overline{\l|H_{v_m^i v_m^j}\r|^2}
\;\;\overline{\l|H_{v_m^{k^\pr} v_m^l}\r|^2} +  \dis\frac{2}{(N_m)^2}\;
\dis\sum_{v_m^i,v_m^j,v_m^{k^\pr},v_m^l} \l[\overline{H_{v_m^i v_m^j}
H_{v_m^{k^\pr} v_m^l}}\r]^2 \nn8
=\l[\Lambda^0(k)\r]^2 + \dis\frac{2}{(N_m)^2}\;\dis\sum_{\nu=0}^{m-k}\;
\l[\Lambda^\nu(k)\r]^2\,d(\nu)
\earr
\ee
Here in the second equality used is the property $\overline{x^2y^2} =
(\overline{x^2})\;(\overline{y^2}) + 2 [\overline{xy}]^2$ of Gaussian
variables $x$ and $y$. Similarly the final result follows by applying (25)
to the second term in the second equality and simplifying the CG
coefficients using Eq. (3). Now, the variance of the  distribution of the
variances of the $H$ spectra over the EGUE($k$)  ensemble is,
\be
{\hat{\Sigma}}_{22} = \dis\frac{\Sigma_{22}}{\l[\overline{\lan
H^2\ran^m}\r]^2}\;-1\;=\;\dis\frac{2}{(N_m)^2}\;\dis\sum_{\nu=0}^{m-k}\;
\l[\dis\frac{\Lambda^\nu(k)}{\Lambda^0(k)}\r]^2 \;d(\nu)
\ee
Eqs. (27,29,31,33) are also given by Benet et al \cite{Be-01}; this paper
neither gives details of the derivations nor uses $SU(N)$ Racah coefficients.
Also in this work, ${\hat{\Sigma}}_{11}$, ${\hat{\Sigma}}_{22}$ and
$\gamma_2+1$ are denoted by $S$, $R$ and $Q$ respectively while we have
followed Ref. \cite{Mo-75}. 

\section{Conclusions}

Going beyond PW, matrix elements of the matrix of second moments are
written explicitly in terms of $SU(N)$ Wigner and Racah coefficients and
this result is obtained by recognizing that the reduced matrix elements of
$B_k(g_\nu)$ are $SU(N)$ Racah coefficients. With this one has Eq. (19)
and this is converted into the generalized eigenvalue expansion form by
first applying a $SU(N)$ Racah transform  and then applying the
Biedenharn-Elliott sum rule extended to $SU(N)$. This gives the eigenvalues
of the matrix of second moments explicitly in terms of $SU(N)$ Racah
coefficients (Eq. (24)). The two different forms given by Eqs. (19,25) for
the covariances of $m$-particle $H$ matrix elements, give in a simple
manner the formulas for the low order moments $M_p$ that define the state
density and the bivariate moments $\Sigma_{pq}$ that give information about
fluctuations. 

Although EGUE($k$) for only fermions is considered in this paper, all the
results in fact translate to those of EGUE($k$) for bosons by using the well
known $N \rightarrow -N$ symmetry \cite{No-75,Ko-80}, i.e. in the fermion
results replace $N$ by $-N$ and then take the absolute value of the final
result. For example, the $m$ boson space dimension is $d(m) = \l|{\tiny{
\l(\barr{c} -N \\ m \earr \r)}}\r| = \{(N-m+1)_m\}$. More importantly the 
eigenvalues of the matrix of the second moments are
\be
\Lambda^\nu_B(k) \rightarrow \l|{\tiny{\l(\barr{c} m-\nu \\ k \earr \r)}}\;
{\tiny{\l(\barr{c} -N-m+k-\nu \\ k \earr\r)}}\r|= {\tiny{\l(\barr{c} m-\nu \\ 
k \earr \r)\;\l(\barr{c} N+m+\nu-1\\ k \earr\r)}}
\ee
This result  was explicitly derived in \cite{As-02}; see Eq. (14) of this
paper. Moreover for bosons, $\{k\} \otimes \{k^{N-1}\} \rightarrow
g_\nu=\{2\nu,\nu^{N-2}\}\;,\;\;\nu=0,1,\cdots,k$. Also, the $N
\rightarrow -N$ symmetry and Eq. (16) give $d(g_\nu) =
\{(N+\nu-1)_\nu\}^2-\{(N+\nu-2)_{\nu-1}\}^2$ and this is same as Eq. (15)
of \cite{As-02}. Similarly Eqs. (27,29,31,33) extend directly to the boson
EGUE($k$) with $\Lambda^\nu(k)$ replaced by $\Lambda^\nu_B(k)$ defined in
Eq. (34). In addition, for fermions to bosons there is also a $m
\leftrightarrow N$ symmetry and this connects fermion results (say for
$M_p$ and $\Sigma_{pq}$) in dilute limit to boson results in dense  limit
\cite{Ko-80}. 

Recently there is considerable interest in mesoscopic physics to study
EGUE($k$) for fermions with spin \cite{Ph-01} and here the embedding algebra 
is $U(2N) \supset U(N) \otimes SU(2)$ with $SU(2)$ generating spin. The
approach presented in Sections II-IV is being applied to this system; some
useful results for the $U(2N) \supset U(N) \otimes SU(2)$  Wigner-Racah
algebra are available in Refs.  \cite{He-74,Kk-01}.  Finally, Wigner-Racah
algebra analysis of embedded  ensembles with more general group symmetries
(see \cite{Ko-01,Pl-02,Kk-01,Pa-04} for examples) should be possible in
future, thus opening up a new direction in random matrix theory.

\acknowledgments
 
Thanks are due to H.A. Weidenm\"{u}ller for reading the first draft of the
paper and for his valuable comments.

\ed